\documentclass[twocolumn,aps,showpacs]{revtex4}
\usepackage{}

\usepackage{amsfonts}
\usepackage{amsmath}
\usepackage{amssymb}
\usepackage{graphicx}

\begin{document}

\title{Kosterlitz-Thouless transition in
disordered two-dimensional topological insulators}
\author{Zhong Xu}
\author{L. Sheng}
\email{shengli@nju.edu.cn}
\author{R. Shen}
\author{Baigeng Wang}
\author{D.Y. Xing}
\email{dyxing@nju.edu.cn} \affiliation{$^1$National Laboratory of
Solid State Microstructures and Department of Physics, Nanjing
University, Nanjing 210093, China\\}
\date{\today }

\begin{abstract}
The disorder-driven metal-insulator transition in the quantum spin Hall 
systems is studied by scaling analysis of the Thouless
conductance $g$. Below a critical disorder strength, the conductance
is independent of the sample size $M$, an indication of critically
delocalized electron states. The calculated beta function
$\beta=d\ln g/d\ln M$  indicates that the metal-insulator transition
is Kosterlitz-Thouless (KT) type, which is characterized by bounding and
unbounding of vortex-antivortex pairs of the local currents. The 
KT like metal-insulator transition is a basic characteristic of the quantum
spin Hall state, being independent of 
the time-reversal symmetry.
\end{abstract}


\maketitle

The Kosterlitz-Thouless (KT) transition, which was first proposed by
Kosterlitz and Thouless in two-dimensional (2D) XY model,~\cite{KT1}
is an important phenomenon in 2D systems, such as 2D magnets,
superconductors and superfluids in thin films. It is a typical
topological phase transition, which has been understood as bounding
and unbounding of vortex-antivortex pairs.~\cite{KT1,KT3}
In recent years, the KT like transition was also found in 2D
disordered electronic models, including the model with random
magnetic fluxes,~\cite{Ktfordisorder1,Ktfordisorder2} and the
graphene model with long-range impurity scattering
potential.~\cite{Xie} These systems show localization-delocalization
transitions with decreasing disorder strength. On the metallic side,
the Thouless conductance $g$ is
independent of the
system size $M$, leading to a vanishing beta function~\cite{Scaling,beta}
defined as $\beta=d\ln g/d\ln M$, which is an essential
characteristic of the KT
transition.~\cite{Ktfordisorder1,Ktfordisorder2,Xie} 
From the
classification scheme based on the universality
classes,~\cite{Review1} the former
model~\cite{Ktfordisorder1,Ktfordisorder2} belongs to the unitary
class, and the existence of delocalized states is due to the random
phases in the hopping integrals, which 
result in the Chern number fluctuations.~\cite{DNShengRF} 
The latter model~\cite{Xie} belongs to the orthogonal
class, and the electron delocalization arises from the nonzero Berry
phases at two Dirac points, which cannot annihilate each other when
the impurity scattering is long-range correlated.

A topological insulator~\cite{KaneR,ZhangR2} is a band insulator
with gapless edge states or surface states traversing the band gap,
which originate from the nontrivial band topology. The gapless
nature of the edge states or surface states is protected by the
time-reversal symmetry. The 2D version of the topological
insulators, which is also called the quantum spin Hall (QSH) state,
was predicted by Kane and Mele~\cite{Pre1} to exist in a graphene
model with intrinsic spin-orbit coupling. Later it was
experimentally realized in mercury telluride (HgTe) quantum
wells.~\cite{Exp1} The robustness of the QSH state is associated
with new types of topological invariants of the bulk energy
bands.~\cite{Z2,Spinchern,cz,Prodan1}

Since topological invariants are carried by bulk
extended states, there have being strong interest in the delocalization
problem in the QSH systems. Onoda $et$ $al.$~\cite{Nagaosa} 
studied the Kane-Mele model using
the transfer matrix method and showed the existence of
extended states in a finite energy region in the presence of time
reversal symmetry. Based upon calculations of the critical exponents, they
suggested that the QSH state belongs to a new universality class
different from the ordinary symplectic universality
class.~\cite{symplectic} 
Consistent result was obtained by Prodan using the level
statistics analysis.~\cite{ProdanLevel} 
Employing a continuum model, 
Mong $et$ $al.$~\cite{Mong} showed that the
surface states of a weak topological insulator with even number of
Dirac cores are delocalized in the absence of a mass term, and
localization occurs when the mass term and strong disorder are
present. This conclusion may also be applicable to the QSH systems.  Xu
$et$ $al.$~\cite{Xu} demonstrated numerically that the metallic
state in the QSH systems remains to be robust even when the
time-reversal symmetry is broken, in glaring contrast to
trivial unitary class, where all electron states are localized. 
This result was attributed to the fact that
the nontrivial band topology, as described 
by the spin Chern numbers, 
is intact when the time-reversal
symmetry is broken.~\cite{Prodan1,Yang}
The existence of delocalized
states in the QSH systems is affirmative, but the basic characteristics 
of the metal-insulator transition have not been well established.


In this Letter, from calculations of the Thouless conductance in the
Kane-Mele model, it is shown that the disorder-driven
metal-insulator transition in the QSH system is a KT phase
transition. Below a critical disorder strength, the Thouless
conductance $g$ is found to be independent of the sample size, an
indication of critically delocalized electron states. The beta
function obtained from the universal scaling
function of the conductance, vanishes in the metallic phase. Through
mapping the local currents onto the local spins in the 2D $XY$
model, we further show that the KT transition is characterized by
bounding and unbounding of vortex-antivortex pairs. The 
KT like metal-insulator transition is a 
fundamental characteristic of the QSH state, 
allowing to be examined experimentally,
which neither depends
on the time-reversal symmetry, nor requires any long-range
correlations of the disorder.


We start from the Kane-Mele model defined on a 2D honeycomb
lattice,~\cite{Pre1,Spinchern,disorder} with the Hamiltonian given
by
\begin{eqnarray}
\label{kanemelemodel}
 H =  - t\sum\limits_{\langle ij\rangle} {c_i^\dag
{c_j}}+\frac{{2i}}{{\sqrt 3 }}{V_{\mbox{\tiny SO}}}\sum\limits_{\langle\langle ij \rangle\rangle }
{c_i^\dag\overset{\lower0.5em\hbox{$\smash{\scriptscriptstyle\rightharpoonup}$}}{\sigma}
\cdot({{\overset{\lower0.5em\hbox{$\smash{\scriptscriptstyle\rightharpoonup}$}}{d}}_{kj}}
\times{{\overset{\lower0.5em\hbox{$\smash{\scriptscriptstyle\rightharpoonup}$}}{d}
}_{ik}}){c_j}}\nonumber \\
+i{V_{\mbox{\tiny R}}}\sum\limits_{\langle ij\rangle}{c_i^\dag\hat{\mbox{e}}_z
\cdot(\overset{\lower0.5em\hbox{$\smash{\scriptscriptstyle\rightharpoonup}$}}{\sigma}
\times{{\overset{\lower0.5em\hbox{$\smash{\scriptscriptstyle\rightharpoonup}$}}{d}}_{ij}}){c_j}}
+\sum\limits_i {{w_i} {c_i}^\dag}  {c_i}\ .
\label{eq1}
\end{eqnarray}
Here, the first term is the usual nearest neighbor hopping term with
$c_i^\dag = (c_{i \uparrow }^\dag ,c_{i \downarrow }^\dag )$ as the
electron creation operator on site $i$. The
second term is the intrinsic spin-orbit coupling with coupling
strength $V_{\mbox{\tiny SO}}$, where $\vec{\sigma}$ are the Pauli
matrices, $i$ and $j$ are two next nearest neighbor sites, $k$ is
their unique common nearest neighbor, and vector $\vec{d}_{ik}$
points from $k$ to $i$. The third term stands for the Rashba
spin-orbit coupling with coupling strength $V_{\mbox{\tiny R}}$.
${w_i}$ is a random on-site energy potential uniformly distributed
between $\left[- W/2 , W/2\right]$, which accounts for nonmagnetic
disorder. For convenience, we will set $\hbar$, $t$ and the distance
between the nearest neighbor sites all to be unity. In the following
calculations, $V_{\mbox{\tiny SO}}=V_{\mbox{\tiny R}}=0.1$ is fixed,
so that the model is inside the QSH phase at $W=0$.

The localization property of an infinite 2D system can be
extracted from the scaling analysis of the dimensionless Thouless
conductance calculated for finite-size samples.~\cite{Scaling,beta}
We consider a rectangular sample with two zigzag sides and two
armchair sides. Each zigzag side has $M$ atoms and each armchair
side has $2M$ atoms, so that the sample has totally $M\times M$
atoms. The rectangular sample is connected to two semi-infinite
leads of the same width with armchair interfaces. At
zero temperature, the two-terminal dimensionless Landauer
conductance $g_{\mbox{\tiny L}}$ is calculated from the
Landauer-B\"uttiker formula~\cite{Landauer} ${g_{\mbox{\tiny L}}}
=\mbox{Tr}[\Gamma_{\mbox{\tiny L}}G^{r}\Gamma_{\mbox{\tiny
R}}G^{a}]$, where $G^{r(a)}(E)$ is the retarded (advanced)
Green's function and $\Gamma_{\mbox{\tiny L(R)}}(E)$ is the coupling
matrix between the sample and the left (right) lead. The Thouless
conductance $g$ is related to the Landauer conductance $g_L$ through
the relation $1/g = 1/{g_{\mbox{\tiny L}}} - 1/{N_{\mbox{\tiny
C}}}$,~\cite{Xie,Braun} where $1/N_{\mbox{\tiny C}}$ is the contact
resistance between the leads and the sample with $N_{\mbox{\tiny C}}$ as the number
of propagating channels in the leads at the Fermi energy $E$.

\begin{figure}[tbp]
\includegraphics[width=2.8in]{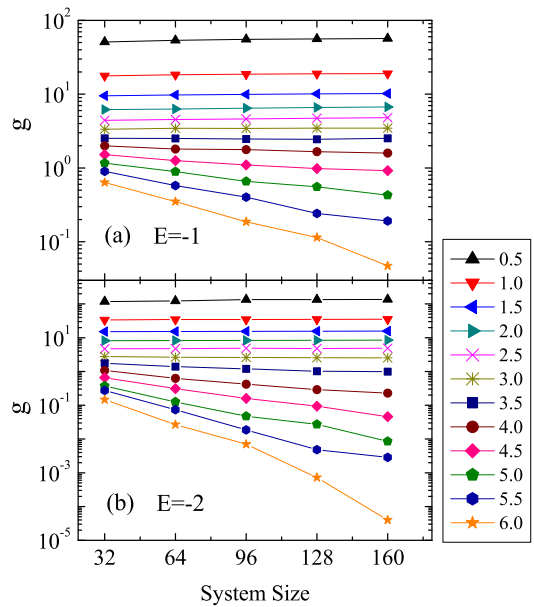}
\caption{(Color online) Conductance as a function of system size $M$ at
$E=-1$ (a) and $E=-2$ (b). Symbols with different colors and
shapes correspond to different disorder strengths $W$. }
\label{Fig.1}
\end{figure}

Figure 1 shows the size dependence of the calculated conductance $g$
in logarithmic scale for different disorder strengths $W$ at $E=-1$
and $-2$. Periodic boundary conditions are employed in the
transverse direction, so that no contribution from edge states is
involved. At given values of $W$ and $M$, the conductance is
averaged over 200 to 500 random disorder configurations. Symbols
with different shapes are used to distinguish data for different
$W$. It is found that the conductance behaves differently with
changing sample size in weak and strong disorder regions. For weak
disorder, $g$ neither decreases nor increases with increasing
sample size. As a result, the conductance is expected to remain
constant in the thermodynamic limit, suggesting that the electron
states  at the Fermi energy are critically delocalized. For strong
disorder, $g$ decreases with increasing sample size,
corresponding to localized states. A metal-insulator transition
occurs at an intermediate critical disorder strength.


\begin{figure}[tbp]
\includegraphics[width=3.3in]{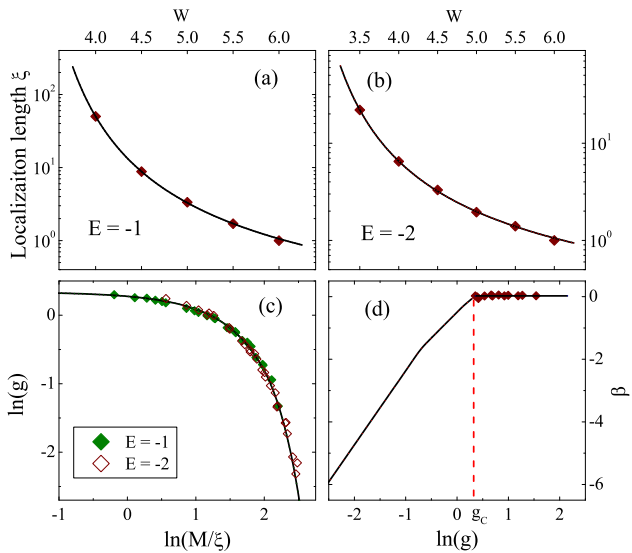}
\caption{(Color online) (a,b) Calculated localization length $\xi$
(symbols), fitted with an exponential function (solid lines). (c)
Universal scaling function $g=f(M/\xi)$, where solid and open
symbols are conductance at $E=-1$ and $-2$, respectively, and the
solid line represents a smooth functional fit to the data. (d)
Calculated beta function.} \label{Fig.2}
\end{figure}

To further investigate the nature of the metal-insulator phase
transition, we perform scaling analysis on the data shown in Fig.\ 1
and calculate the beta function $\beta=d\ln g/d\ln M$. According to
the one-parameter scaling theory,~\cite{Scaling,beta} the
localization length in the thermodynamic limit can be extracted by
fitting all the data on the insulator side in Fig.\ 1 into a
universal function $g=f(M/\xi)$, where the fitting parameter $\xi$
yields the localization length in the thermodynamic limit. The
obtained localization lengths for $E=-1$ and $-2$  are plotted in
Figs.\ 2a and 2b, respectively, and the universal scaling function
$g=f(M/\xi)$ is shown in Fig.\ 2c.
In the above scaling procedure, only the relative magnitudes of the
localization lengths $\xi$ can be determined. Therefore, we set $\xi$
at $W=6.0$ to be unity, and the localization lengths for other $W$
are the relative values with respect to it. It is found that the
localization length $\xi$ as a function of disorder strength $W$
shown in Figs.\ 2(a,b) can be fitted with an exponential
function $\xi\propto e^{\alpha /\sqrt {W - {W_{\mbox{\tiny C}}}}}$
with $\alpha$ and $W_{\mbox{\tiny C}}$ as two fitting parameters.
This behavior indicates an exponential decay of the localization
length with increasing the disorder strength in the insulating
phase, which is typical of a KT
phase transition.~\cite{KT1,KT3} 
The critical disorder strength for such a metal-insulator transition
is obtained as $W_{\mbox{\tiny C}}=3.12$ at $E=-1$ and 2.54 at
$E=-2$, respectively. To facilitate the calculation of the 
beta function $\beta=d\ln g/d\ln M$, we fit the data in Fig.\ 2c
with a smooth function, as indicated by the solid line, which
is explained below. In the region
of small $\ln(M/\xi)$, the data is fitted with $g =
g_{\mbox{\tiny C}}{e^{-(M/\xi)^{1/\nu }}}$, which is consistent with
the general behavior of the beta function, $\beta  = \frac{1}{\nu
}\ln (g/{g_{\mbox{\tiny C}}})$, in the quantum critical
region.~\cite{DNSheng}
On the other hand, a polynomial fit is used in the region
of relatively large $\ln(M/\xi)$. At the boundary between the
two regions, $\ln g$ and its derivative
$\beta$ are required to be continuous.

After obtaining the smooth fitting function describing the scaling
relation between $\ln g$ and $\ln M$, we take its derivative  to
calculate the beta function $\beta=d\ln g/d\ln M$, as plotted in
Fig.\ 2d. Here, we note that for weak disorder ($W<W_{\mbox{\tiny
C}}$), where the conductance is nearly independent of the sample
size, the localization length 
diverges and cannot be treated with
the same scaling procedure as above. Instead, we fit the logarithmic
conductance $\ln g$ as a function of $\ln M$ with straight lines,
and take their slopes as the values
of $\beta$, which are shown by the diamonds in Fig.\ 2d. 
We see from Fig.\ 2d that $\beta$ increases with $\ln g$, and
vanishes on average above a critical point ($g>g_{\mbox{\tiny C}}\simeq 1.3$)
with negligible fluctuations, indicating clearly that the
disorder-driven metal-insulator transition is a KT transition, as
observed in other disordered 2D electron
systems.~\cite{Ktfordisorder1,Ktfordisorder2,Xie}
The behavior of the beta function 
also suggests that a minimum (critical) conductivity,
$\sigma_{c}=g_{\mbox{\tiny C}}(e^2/h)/\sqrt{3}\simeq 0.75(e^2/h)$ ($1/\sqrt{3}$ being
the aspect ratio), exists
for the bulk QSH system.

\begin{figure}[tbp]
\includegraphics[scale=0.45]{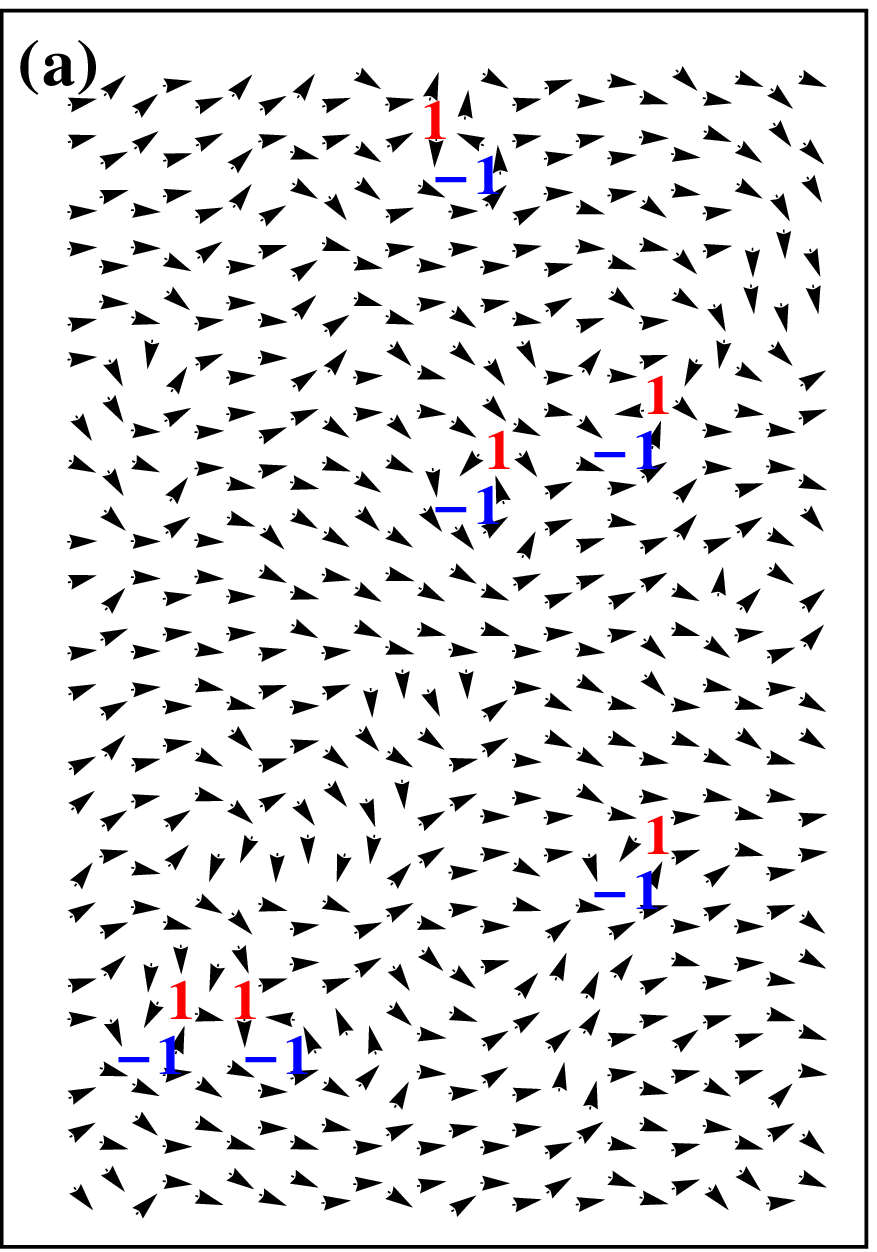}
\includegraphics[scale=0.44]{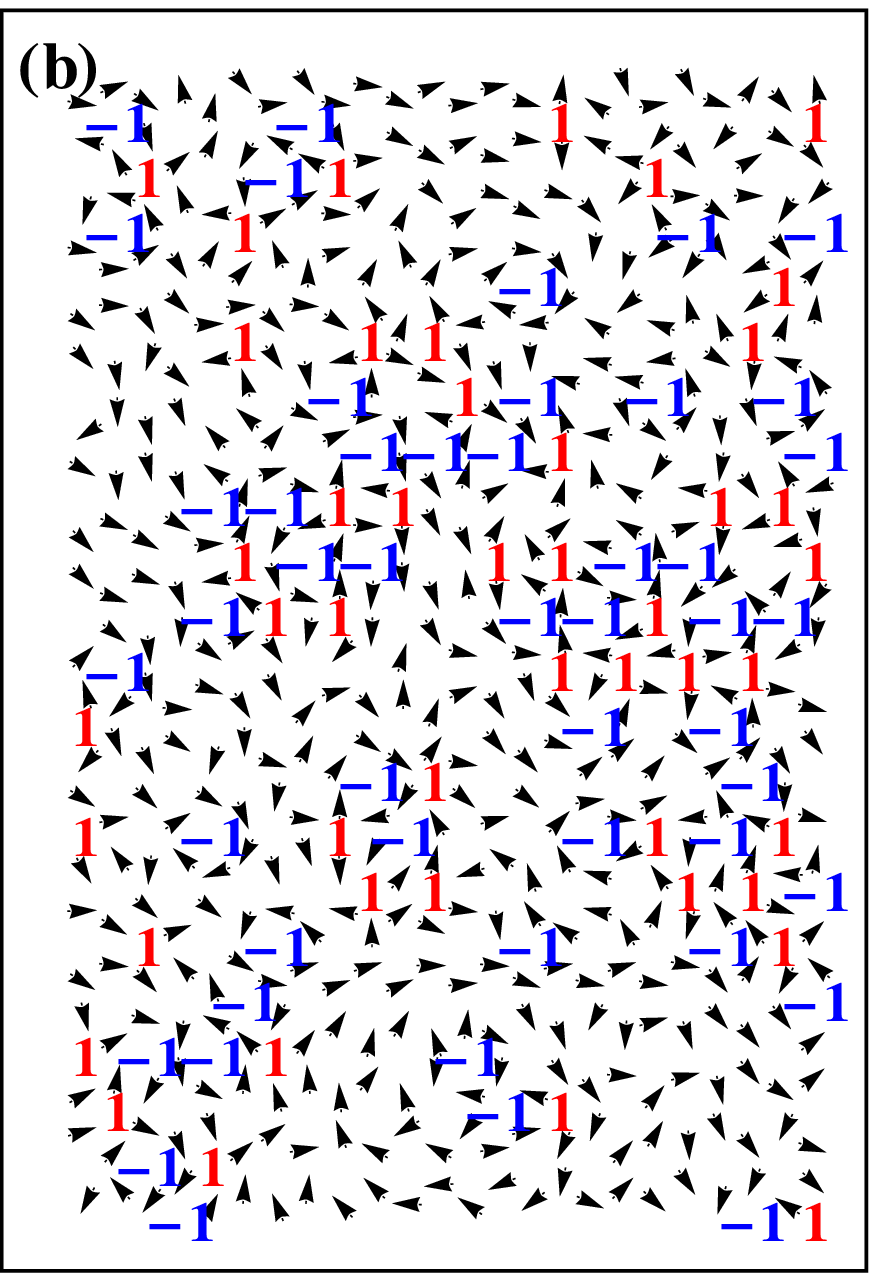}
\caption{(Color online) Typical distribution of vortices (red positive numbers)
and antivortices (blue negative numbers) on
the two sides of the KT like transition
in a sample with $40\times 40$ sites at $E=-2$
for (a) $W=1.5t$ (metallic phase) and (b) $W=3.5t$ (insulator phase).
For clarity, only a part of the sample is shown.}
\label{Fig.3}
\end{figure}
An important characteristic of the traditional KT transition in the
$XY$ model is the  bounding of vortex-antivortex pairs in the
ordered phase near the transition point and unbounding in the
disordered phase.~\cite{KT1,KT3} For 2D electron systems, Zhang $et$
$al.$~\cite{Xie} proposed to map the local currents onto the local
spins in the $XY$ model. The bond current vector between sites $m$
and $n$
can be calculated by~\cite{Xie,localcurrent1,localcurrent3} 
\begin{equation}
{i_{m{\sigma _2} \to n{\sigma _1}}} = \frac{{ - 2e}}{\hbar }\int
{\frac{{dE}}{{2\pi }}{\mathop{\rm Re}\nolimits} [{H_{m{\sigma
_2},n{\sigma _1}}}G_{n{\sigma _1},m{\sigma _2}}^ <(E) ]},
\end{equation}
where ${\sigma _1}$ and ${\sigma_2}$ represent the spin index and
$G_{n{\sigma_1},m{\sigma_2}}^<(E)$ is the matrix element of the
lesser Green's function. The lesser Green's function is given by
\begin{equation}
{G^{<}}(E)={G^{r}}(E)[i\sum\limits_{\alpha }{{\Gamma _{\alpha }}(E)f_{\alpha
}}(E)]{G^{a}}(E)\ ,
\end{equation}%
where ${f_\alpha }(E) = {f_0}(E + e{V_\alpha })$ is the 
Fermi distribution function in lead $\alpha$
with $V_{\alpha}$ as the electrical potential in the lead.
The local current vector defined on site $n$ is ${i_n} =
\sum_{m{\sigma _2}{\sigma _1}} {{i_{m{\sigma _2}
\rightarrow n{\sigma _1}}}}$,~\cite{Xie} where the vectorial
summation is taken over all the nearest and next nearest neighbors
$m$ of site $n$. Along a closed path, the polar angle $\theta_n$ of
$i_n$ is considered to change continuously with changing
coordinates, and a topological charge is defined as
$c=\frac{1}{2\pi}\oint\nabla\theta\cdot dl$.~\cite{Xie} 
A nonzero number of $c$ indicates the appearance 
of a vertex $(c>0)$ or an antivortex $(c<0)$.
Figure 3
shows typical distributions of the topological charges on the two
sides of the metal-insulator transition. The arrows in the figure
indicate only the direction of the local currents, without showing
the magnitude for the purpose of visualization. In the delocalized
phase (Fig.\ 3a), a few vortices ($c>0$) and antivortices ($c<0$)
are excited in pairs and bounded together, corresponding to the
ordered phase at low temperature in the 2D $XY$ model. In the
localized phase  (Fig.\ 3b), a large number of vortices and
antivortices appear and are mostly unbounded, corresponding to the
disordered phase at high temperature.

\begin{figure}[tbp]
\includegraphics[width=3.2in]{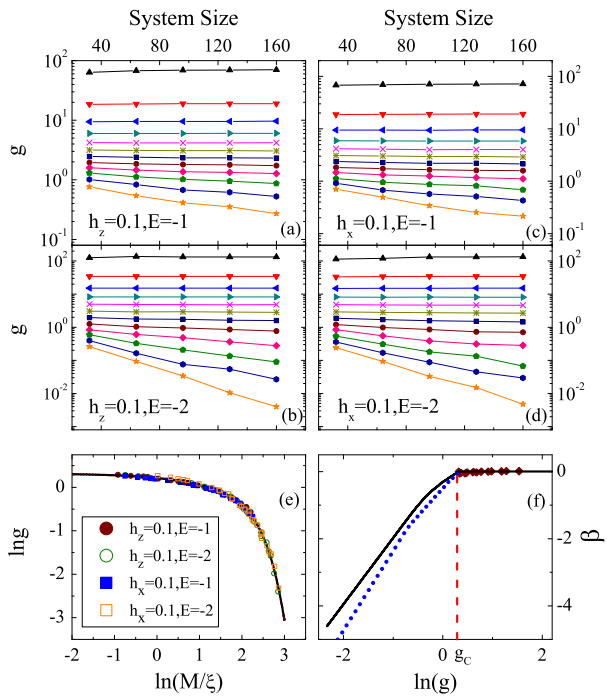}
\caption{(Color online) (a-d) Conductance as a function of sample
size with Fermi energy $E$, in the presence of a vertical Zeeman
field $h_z$ or a horizontal Zeeman field $h_x$. (e) Universal
scaling function $g=f(M/\xi)$, where the solid line represents a
smooth functional fit to the data. (f) The corresponding beta
functions for the time-reversal symmetry broken QSH system (solid
line) and time-reversal symmetry invariant QSH system (dotted
line).} \label{Fig.4}
\end{figure}

Finally, we study the effect of breaking the time-reversal symmetry
on the KT phase transition, by including a new term
$\vec{h}\cdot\vec{\sigma}$ into Hamiltonian (\ref{kanemelemodel}),
where $\vec{h}$  stands for a uniform Zeeman field.
The calculated conductances at $E=-1$ and $-2$ as functions of
sample size $M$ for different disorder strengths  are shown in Fig.\
4(a,b) for a vertical Zeeman field $\vec{h}=(0, 0, h_z =0.1)$  and
in Fig.\ 4(c,d) for a horizontal field $\vec{h}=(h_x =0.1, 0, 0)$.
The parameters are chosen in such a manner that the system is in the
time-reversal symmetry broken QSH phase in the clean
limit.~\cite{Yang,Xu} The conductance shown in Fig.\ 4(a-d) displays 
similar size dependence to that in Fig.\ 1. For weak disorder,
the conductance is nearly independent of sample size, indicating the
existence of the critically delocalized states. For strong disorder,
the conductance decreases with sample size, indicating that the
electron states are localized. It is found that all the data on the
insulating side in Fig.\ 4(a-d) can also be fitted with a universal
scaling function $g=f(M/\xi)$, as shown in Fig.\ 4e. In Fig.\ 4f,
the beta function (solid line) for the time-reversal symmetry broken
QSH system is obtained in the same way as in Fig.\ 3. As expected,
the beta function vanishes above a critical conductance
$g_{\mbox{\tiny C}}$, exhibiting the essential feature of the KT
like transition.
For comparison, $\beta$ for the time-reversal invariant QSH system
obtained in Fig.\ 2d is shown by the dotted line in Fig.\ 4f. Both
curves exhibit the same characteristic of the KT phase transition, and
the critical points $g_{\mbox{\tiny C}}$ almost coincide. However,
the two curves deviate from each other on the insulating side, which
may be attributed to different symmetries.

In summary, the metal-insulator transition in the QSH systems is
studied based upon the scaling analysis of the Thouless conductance.
The disorder-driven metal-insulator transition is found to be a KT
like transition, characterized by bounding and unbounding of
vortex-antivortex pairs of local electrical currents. The beta
functions for the QSH systems with and without the time-reversal
symmetry are close to each other on the metallic side, but somewhat
different on the insulating side, the latter being attributed to the fact
that they belong to different symmetry classes. Our work established
a fundamental property of the QSH state, which can be observed 
experimentally.


\section*{Acknowledgments}
This work was supported by the State Key Program for Basic Researches of
China under Grant Nos. 2009CB929504 (LS), 2011CB922103 (BW) and
2010CB923400 (DYX), the National Natural Science Foundation of China under
Grant Nos. 11074110 (LS), 11074111 (RS), 60825402, 11023002 (BW), 11174125,
11074109 (DYX), and by a project funded by the PAPD of Jiangsu Higher
Education Institutions. 




\begin{thebibliography}{99}
\bibitem{KT1} J. M. Kosterlitz and D. J. Thouless, J. Phys. C \textbf{6}, 1181 (1973);
J. M. Kosterlitz, J. Phys. C \textbf{7}, 1046 (1974).
\bibitem{KT3} Z. Gul{\'a}csi and M. Gul{\'a}csi, Advances in Physics \textbf{47}, 1 (1998).
\bibitem{Ktfordisorder1} S. -C. Zhang and D. P. Arovas, Phys. Rev. Lett. \textbf{72}, 1886 (1994).
\bibitem{Ktfordisorder2}  X. C. Xie, X. R. Wang and D. Z. Liu, Phys. Rev. Lett. \textbf{80}, 3563 (1998).
\bibitem{Xie} Y. Y. Zhang, J. P. Hu, B. A. Bernevig, X. R. Wang, X. C. Xie and W. M. Liu, Phys. Rev. Lett. \textbf{102}, 106401 (2009).
\bibitem{Scaling} A. MacKinnon and B. Kramer, Z. Phys. B \textbf{53}, 1 (1983).
\bibitem{beta} E. Abrahams, P.W. Anderson, D. C. Licciardello, and T.V. Ramakrishnan, Phys. Rev. Lett. \textbf{42}, 673 (1979).
\bibitem{Review1} F. Evers and A. D. Mirlin, Rev. Mod. Phys. \textbf{80}, 1355 (2008).
\bibitem{DNShengRF} D. N. Sheng and Z. Y. Weng, Phys. Rev. Lett. {\bf 75}, 2388 (1995).
\bibitem{KaneR} M. Z. Hasan and C. L. Kane, Rev. Mod. Phys. \textbf{82}, 3045 (2010).
\bibitem{ZhangR2} X. L. Qi and S. C. Zhang, Rev. Mod. Phys. \textbf{83}, 1057 (2011).
\bibitem{Pre1} C. L. Kane and E. J. Mele, Phys. Rev. Lett. \textbf{95}, 226801
(2005).
\bibitem{Exp1}M. Ko\"nig, S. Wiedmann, C. Brune, A. Roth, H. Buhmann, L. W. Molenkamp, X. L. Qi, and S. C. Zhang, Science \textbf{318}, 766
(2007).
\bibitem{Z2}C. L. Kane and E. J. Mele, Phys. Rev. Lett. \textbf{95}, 146802 (2005).
\bibitem{Spinchern}D. N. Sheng, Z. Y. Weng, L. Sheng, and F. D. M. Haldane, Phys. Rev. Lett. \textbf{97}, 036808 (2006).

\bibitem{cz} T. Fukui, and Y. Hatsugai, Phys. Rev. B \textbf{75} 121403 (2007); A. M. Essin, and J. E. Moore, Phys. Rev. B \textbf{76}, 165307 (2007);
\bibitem{Prodan1}E. Prodan, Phys. Rev. B \textbf{80}, 125327 (2009); E. Prodan, New J. Phys. \textbf{12}, 065003 (2010).
\bibitem{Nagaosa} M. Onoda, Y. Avishai and N. Nagaosa, Phys. Rev. Lett. \textbf{98}, 076802 (2007).
\bibitem{symplectic} Y. Asada, K. Slevin, and T. Ohtsuki, Phys. Rev. Lett. \textbf{89}, 256601 (2002).
\bibitem{ProdanLevel} E. Prodan, J. Phys. A: Math. Theor. 44, 113001 (2011).
\bibitem{Mong} R.S.K. Mong, J. H. Bardarson, and J. E. Moore, Phys. Rev. Lett. \textbf{108}, 076804 (2012).

\bibitem{Xu} Z. Xu, L. Sheng, D. Y. Xing, E. Prodan and D. N. Sheng, Phys. Rev. B \textbf{85}, 075115 (2012).
\bibitem{Yang}Y. Yang, Z. Xu, L. Sheng, B. Wang, D. Y. Xing and D. N. Sheng, Phys. Rev. lett. \textbf{107}, 066602 (2011).
\bibitem{disorder}L. Sheng, D. N. Sheng, C. S. Ting, and F.D.M. Haldane, Phys. Rev. Lett. \textbf{95}, 136602
(2005).

\bibitem{Landauer} S. Datta, Electronic Transport in Mesoscopic Systems (Canmbridge University Press, Cambridge, England,
1995).
\bibitem{Braun} D. Braun, E. Hofstetter, G. Montambaux and A. MacKinnon, Phys. Rev. B \textbf{55}, 7557 (1997).
\bibitem{DNSheng} D. N. Sheng and Z. Y. Weng, Phys. Rev. Lett. \textbf{83}, 144 (1999).
\bibitem{localcurrent1} A. Cresti, G. Grosso, and G. P. Parravicini, Phys. Rev. B \textbf{69}, 233313 (2004).
\bibitem{localcurrent3} Y. Xing, L. Zhang, and J. Wang \textbf{84}, 035110 (2011).
\end{thebibliography}
\end{document}